\documentclass[11pt]{article}
\usepackage{latexsym}  
\usepackage{amssymb}
\usepackage{graphicx}
\usepackage{amsmath}

\begin{document}

\begin{center}
{\Large\bf  
  Rare kaon decay  $K^+ \rightarrow \pi^-   \mu^+  \mu^+$
as the key event 

for  the  right-handed weak interaction effects
}\\[0.1in]

Yoshio Koide

 {\it ${}^a$ Department of Physics, Osaka University, 
Toyonaka, Osaka 560-0043, Japan} \\
{\it E-mail address: koide@epp.phys.sci.osaka-u.ac.jp}

\end{center}

\begin{quotation} 
We discuss on the search for the right-handed weak interaction 
effects in the SU(2)$_L \times$SU(2)$_R$ model with
lepton doublets $(\nu_\ell, \ell^-)_L$ 
and $(N_\ell, \ell^-)_R$ ($\ell = e, \mu, \tau)$. 
We will point out that only the chance of the observation 
of the  right-handed weak interaction effect will be 
in the rare decay $K^+ \rightarrow \pi^-   \mu^+  \mu^+$. 
\end{quotation}

\vspace{5mm}

{\large\bf 1. \ Introduction} \ 

\vspace{2mm}

We are interested in the right-handed weak interaction effect  
in the SU(2)$_L \times$SU(2)$_R\times U(1)_{B-L}$ 
model \cite{L-Rmodel}. 
Let us denote the lepton  doublets as
$$
\left(
\begin{array}{c}
 \nu_\ell \\
 \ell^- \\
\end{array}
\right)_L , \ \ \ 
\left(
\begin{array}{c}
 N_\ell \\
 \ell^- \\
\end{array}
\right)_R, \ \ \ \ \ (\ell = e, \mu. \tau). 
\eqno(1.1)
$$
Here, the neutral leptons $\nu_\ell$ and $N_\ell$ have 
Majorana masses separately.

Of course, the right-handed lepton doublets 
$(N_\ell, \ell^-)_L$
interact with the right-handed weak boson $W_R$. 
At present, it is known \cite{WRmass} that 
the mass of $W_R$ is  experimentally   
$$
M(W_R) > 3 \  {\rm TeV}. 
\eqno(1.2)
$$ 
Therefore, events induced by the right-handed weak 
interaction are highly suppressed by the factor 
$$
\left( \frac{M(W_L)}{M(W_R)} \right)^4 < 
\left( \frac{ 0.1\ {\rm TeV} }{ 4 \ {\rm TeV}}  \right)^4 
\sim 10^{-6} .
\eqno(1.3)
$$
Here and hereafter, for simplicity, we take $g_L = g_R$ 
for the gauge coupling constants with the weak bosons
$W_L$ and $W_R$. 

For example, we know that the $Ds$ meson 
(mass is 1.9684 GeV) decays into muon 
and neutrino with the branching ratio 
$Br(Ds \rightarrow \mu + \nu_\mu) = 5,50 \pm
0.23$ \%  \cite{PDG} . 
If  there is the right-handed weak boson $W_R$, 
the leptonic  decay  is also possible by mediating 
$W_R$. 

In general, the momenta $p_1$ and $p_2$ in the final state at the rest frame
of parent particle are given by
$$
p_1 = p_2 = \frac{1}{M} \left[ M^4 -2M^2 (m_1^2 +m_2^2)
+ m_1^4 +m_2^4 - 2 m_1^2 m_2^2 \right]^{1/2}.
\eqno(1.4)
$$
Here, $M$ is a mass of the ps-meson in the initial state, 
and $m_1$ and $m_2$ are lepton masses of the final state.
Therefore, the momenta of the final state muon are 
quite different from each other:

{\bf Event via $W_L$}
$$
(p_\mu)_L \simeq 
 \frac{1}{2M} \left( M^4 -2M^2 m_\mu^2 +m_\mu^4 \right)^{1/2} 
= \frac{M}{2}\left(1-\frac{m_\mu^2}{M^2} \right) ,  
\eqno(1.5)
$$

{\bf Event via $W_R$}
$$
(p_\mu)_R \simeq \frac{1}{2M} \left( M^4 -2M^2 m_N^2 +m_N^2 \right)^{1/2}
= \frac{M}{2} \left( 1 -\frac{m_N^2}{M^2} \right) , 
\eqno(1.6)
$$
where $M$ is the mass of $Ds$ meson.
Thus, by measuring the momentum of the muon, i.e. (1.5) or (1.6), 
we can know that the case is due to $W_L$ or $W_R$.  

However, it is noted that the events (1.6) are highly suppressed
by the factor (1.3), so that we  cannot  observe the events 
 substantially.  
 
In this paper, we would like point out that only the visible 
effect induced by $W_R$ is a rare decay 
$K^+ \rightarrow \pi^-   \mu^+  \mu^+$. 

\vspace{5mm}
{\large\bf 2. Rare decay $K^+ \rightarrow \pi^-   \mu^+  \mu^+$}

The rare decay $K^+ \rightarrow \pi^-   \mu^+  \mu^+$ is only 
possible event which we can observe under the right-handed 
weak interaction. 
In this section, we will discus the detail.  

For convenience, let us give only the leptonic part of the 
matrix element.  
As  we show in Appendix A, 
the matrix element of the lepton part 
due to the exchange of $W_L$ for  
$K^+ \rightarrow \pi^-   \mu^+  \mu^+$ is given by 
$$
{\cal M}_L = m_\nu \overline{v^c}(k_1) q\!\! /  
 \frac{i C^{-1}}{k^2 -m_\nu^2}
 \frac{1-\gamma_5}{2}p\!\! / v(k_2) 
\eqno(2.1)
$$
and that  due to the exchange of $W_R$ is 
given by 
$$
{\cal M}_R =  M_N \overline{v^c}(k_1) q\!\! /  
 \frac{i C^{-1}}{k^2 -M_N^2}
 \frac{1+\gamma_5}{2}p\!\! / v(k_2) .
\eqno(2.2)
$$

Then, for the case  $M_N^2 \ll k^2$, we get

[Case A]
$$
R_{R/L}^A \equiv \frac{{\cal M}_R}{{\cal M}_L} 
\sim \frac{M_N/k^2}{m_\nu/k^2} = \frac{M_N}{m_\nu}.
\eqno(2.3)
$$

On the other case, for the case $M_N^2 \gg k^2$, we get

[Case B]
$$
R_{R/L}^B \equiv \frac{{\cal M}_R}{{\cal M}_L} 
\sim \frac{1/M_N}{m_\nu/k^2} = \frac{k^2}{m_\nu M_N} . 
\eqno(2.4)
$$
Note that the case $M_N^2 \gg k^2$ does not always give 
a large enhancement of $R_{R/L}$ as we see in Eq.(2.4).

Let us see the numerical value of the ratio  $R_{R/L}$. 
For example, we take $m_{\nu_\mu} \sim 10 \ {\rm meV} 
\sim 10^{-11}$ GeV, 
and  $k^2 \sim 0.2$\ GeV$^2$.

Since we want as possible as large value of $R_{R/L}$, 
for the case A, we take $M_N \sim 10^{-1} $ \ GeV. 
Then, we obtain 

[Case A]
$$
R_{R/L}^A \sim 
  \frac{M_N}{m_\nu}  \sim \frac{10^{ -1} }{10^{-11}} \sim 10^{10} .
\eqno(2.5)
$$
On the other hand, for the case B, for the $M_N$ mass
$M_N \sim 10 $\ GeV, we obtain

[Case B]
$$
R_{R/L}^B \sim  \frac{k^2}{m_\nu M_N} \sim 
 \frac{10 \ {\rm GeV}^2 }{10^{-11} \ {\rm GeV} 
\   10 \ {\rm GeV} } \sim 10^{11}  .  
\eqno(2.6)
$$

Of course, since the ratio $R_{R/L}$ is the ratio for the 
lepton part, so that we must multiply the weak boson mass 
ratio $(M_{WL}/M_{WR})^4 \sim 10^{-6}$. 
$$
\left( \frac{M(W_L)}{M(W_R)} \right)^4 R_{R/L}^A \sim 10^4 ,
\ \ \ \ \ \ \ 
\left( \frac{M(W_L)}{M(W_R)} \right)^4 R_{R/L}^B \sim 10^5 .
\eqno(2.7)
$$
(Note that this diagram has two weak boson exchanges. 
But the ratios (2.3) and (2.4) are ones for the amplitude.)
Therefore, our result (2.7) shows that 
the right-handed weak interaction effect is enhanced 
in the rare decay $K^- \rightarrow \pi^- \mu^+ \mu^+$. 
We can confirm by seeing whether the muons in the final state
are $\mu_R$  not $\mu_L$.


\vspace{5mm}

{\large\bf 4. \ Conclusion } 

In conclusion, 
only the chance of the search for the right-handed weak 
interaction will be in the rare decay 
$K^+ \rightarrow \pi^- \mu^+ \mu^+$. 
If we observe the decay  $K^+ \rightarrow \pi^- \mu^+ \mu^+$
in a near future, it will be caused by the exchange of 
the right-handed weak boson. 
This will be confirmed by the helicity of muons in the final state,
i.e. $(\mu^+)_R$.

\vspace{2mm}

%
%

{\large\bf Acknowledgements} 

\vspace{2mm}

The author  is grateful to Prof. M. Tanaka 
for his enjoyable discussion and his helpful comments. 
And, also, the author is grateful to  Prof. T. Yamashita 
for his helpful assistance in preparing this manuscript.
And he is also grateful to Prof.  A. Sato for his 
valuable advice on the experimental behavior of muon.  
This work is supported by JPS KAKENHI Grant 
number JP1903826.

\vspace{15mm}

{\large\bf Appendix A \ 
-- \  Decay $K^+ \rightarrow \pi^- \mu^+ \mu^+$ }\ 

There are two diagrams for the $K^+ \rightarrow 
\pi^- \mu^+ \mu^+$ decay as thown in Fig.1 snd Fig.2. 

\begin{figure}[h]

\vspace{-4mm}
\hspace{20mm}
\begin{picture}(200,150)(0,0)
  \includegraphics[height=.26\textheight]{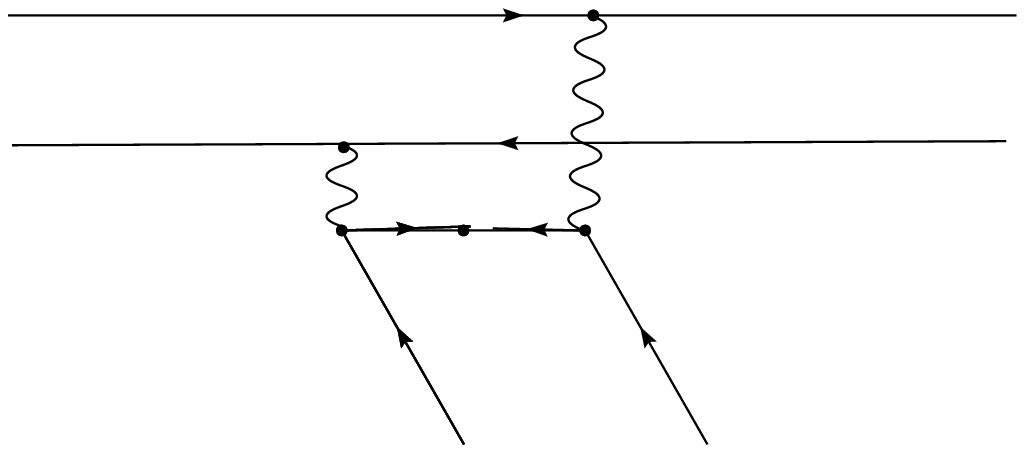}
\end{picture} 

\vspace{-41mm}
\hspace{33mm}$\overline{s}$  \vspace{0mm}\hspace{95mm}$\overline{u}$

\vspace{5mm}\hspace{33mm}$u$    \vspace{0mm}\hspace{95mm}$d$

\vspace{6mm}\hspace{73mm}$\nu$ \vspace{0mm}\hspace{5mm}$\bar{\nu}$

\vspace{12mm}\hspace{76mm}$\mu^+(k_1)$    
\vspace{0mm}\hspace{10mm}$\mu^+(k_2)$

\end{figure}
\vspace{3mm}\hspace{20mm} 
{Fig.1 \ \ Box-type diagram for $K^+(q) \rightarrow \pi^-(p)
 + \mu^+(k_1) + \mu^+(k_2)$ decay}
 \label{fig1}


\begin{figure}[h]
\hspace{25mm}

\hspace{30mm}
\begin{picture}(-180,125)(0,0)
  \includegraphics[height=.26\textheight]{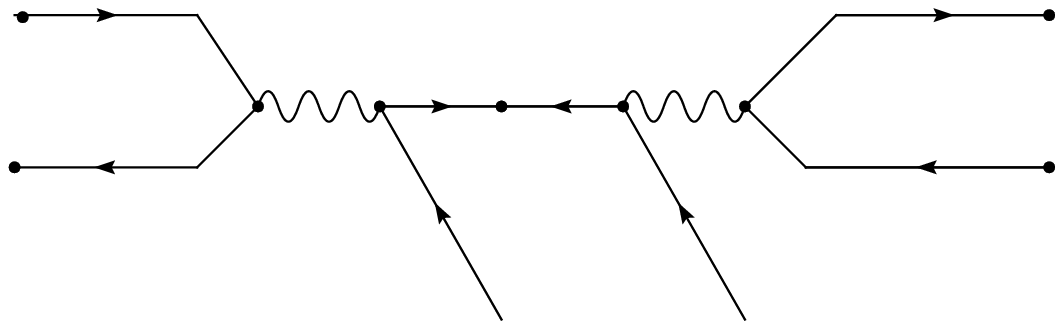}
\end{picture} 

\vspace{-32mm}\hspace{40mm}$\bar{s}$    \vspace{0mm}\hspace{98mm}$\bar{u}$

\vspace{7mm}\hspace{40mm}$u$    \vspace{0mm}\hspace{98mm}$d$

\vspace{-7mm}\hspace{85mm}$\nu$ \vspace{0mm}\hspace{5mm}$\bar{\nu}$

\vspace{15mm}\hspace{88mm}$\mu^+(k_1)$  \vspace{0mm}\hspace{10mm}$\mu^+(k_2)$

\end{figure}

\vspace{0mm}\hspace{20mm} 
{Fig.2 \ \ Tree-type diagram for $K^+(q) \rightarrow \pi^-(p)
 + \mu^+(k_1) + \mu^+(k_2)$ decay}

\vspace{5mm}

However, our interest is only in the lepton part whose 
matrix element is given by the following:
$$
{\cal M}_L \equiv q_\nu p_\mu
\, \overline{v^c}(k_1) \gamma^\nu \frac{1-\gamma_5}{2} 
\left( \frac{i C^{-1} (k\!\! / +m_\nu )}{k^2 -m_\nu^2 +i\varepsilon} 
\right)
\gamma^\mu \frac{1+\gamma_5}{2} v(k_2)
$$
$$
=\overline{v^c}(k_1) q\!\! / \frac{1-\gamma_5}{2} 
 \frac{i C^{-1}}{k^2 -m_\nu^2}(k\!\! / +m_\nu )
p\!\! / \frac{1+\gamma_5}{2} v(k_2) 
$$
$$
=\overline{v^c}(k_1) q\!\! /  
 \frac{i C^{-1}}{k^2 -m_\nu^2}
\frac{1-\gamma_5}{2}(k\!\! / +m_\nu )
 \frac{1-\gamma_5}{2}p\!\! / v(k_2)
$$ 
$$
= m_\nu \overline{v^c}(k_1) q\!\! /  
 \frac{i C^{-1}}{k^2 -m_\nu^2}
 \frac{1-\gamma_5}{2}p\!\! / v(k_2) 
\eqno(A.1)
$$
Here, we have used the following formulae
$$
\Psi^c \equiv C \bar{\Psi}^T ,
\eqno(A.2)
$$
$$
\bar{\Psi}^c = (\Psi^c)^\dagger \gamma^0 =
(C \bar{\Psi}^T)^\dagger \gamma^0 = \Psi^T C (\gamma^0)^2
= - \Psi^T C ,
\eqno(A.3)
$$
$$
C^{-1} \gamma^\mu C = - (\gamma^\mu)^T ,
\eqno(A.4)
$$
$$
C^{-1} \gamma^5 C = \gamma^5 ,
\eqno(A.5)
$$
$$
C= C^{-1} \ \ \ \ \ (C^2={\bf 1} )  . 
\eqno(A.6)
$$
The propagator of Majorana neutrino was quoted from
the Haber and Kane's paper \cite{HK1985}.

Similarly to the case of the $W_L$ exchange, (A.1), 
the result in the $W_R$ exchange is 
given as follows:
$$
{\cal M}_R =  M_N \overline{v^c}(k_1) q\!\! /  
 \frac{i C^{-1}}{k^2 -M_N^2}
 \frac{1+\gamma_5}{2}p\!\! / v(k_2) .
\eqno(A.7)
$$

\vspace{5mm}

\vspace{15mm}



\end{document}